# Inclusive photoproduction of $D^{*\pm}$ mesons at next-to-leading order in the General-Mass Variable-Flavor-Number Scheme

B. A. Kniehl[1], G. Kramer[1], I. Schienbein[2] and H. Spiesberger[3]

[1] II. Institut für Theoretische Physik, Universität Hamburg,
Luruper Chaussee 149, D–22761 Hamburg, Germany

[2] Laboratoire de Physique Subatomique et de Cosmology,
Universite Joseph Fourier, CNRS/IN2P3, INPG,
53 Avenue des Martyrs, F–38026 Grenoble, France

[3] Institut für Physik, Johannes-Gutenberg-Universität,
Staudinger Weg 7, D–55099 Mainz, Germany

**Abstract**

We discuss the inclusive production of $D^{*\pm}$ mesons in $\gamma p$ collisions at DESY HERA, based on a calculation at next-to-leading order in the general-mass variable-flavor-number scheme. In this approach, $\overline{\text{MS}}$ subtraction is applied in such a way that large logarithmic corrections are resummed in universal parton distribution and fragmentation functions and finite mass terms are taken into account. We present detailed numerical results for a comparison with data obtained at HERA and discuss various sources of theoretical uncertainties.

# 1 Introduction

During the last ten years, the two HERA collaborations, H1 and ZEUS, performed several measurements of inclusive $D^{*\pm}$ meson production in photon-proton collisions (see Refs. [1, 2] for the most recent measurements and Refs. [3, 4] for previous ones). Photoproduction in $ep$ collisions at HERA is characterized by an almost vanishing virtuality ($Q^2 \simeq 0$) of the exchanged photon.

Theoretically, the description of photoproduction of charm quarks which fragment into an observed $D^{*+}$ meson, is complicated due to the fact that two interaction modes contribute: the direct process, where the photon interacts directly with a parton originating from the proton, and the resolved process, where the photon acts as a source of partons, which interact with partons in the proton. At next-to-leading order (NLO) in perturbative QCD, these two processes are interrelated. In addition, theoretical predictions are technically difficult to obtain due to the presence of two different scales in the process. On the one hand, the charm-quark mass $m$ can be considered as the large scale, since $m > \Lambda_{\text{QCD}}$, making perturbative QCD applicable. On the other hand, if the transverse momentum $p_T$ of the produced heavy quark is large compared with the heavy-quark mass, $p_T \gg m$, then $p_T$ acts as the dominant large scale for the perturbative calculation. NLO results corresponding to the first situation are reliable, when $m$ is the only large scale, as, for example, in calculations of the total cross section or of the $p_T$ distribution as long as $p_T$ is not much larger than $m$. However, when $p_T \gg m$, large logarithms of the type $\ln(p_T^2/m^2)$ arise to all orders, so that fixed-order perturbation theory is no longer valid. These logarithms can be resummed to improve the perturbative series. Thus, depending on whether $p_T \gg m$ or $p_T \leq m$, different calculational schemes have to be applied.

For $p_T \leq m$, the so-called fixed-flavor-number scheme (FFNS) [5] is applied, where one assumes that the gluon and the light quarks $(u, d, s)$ are the only active partons within the proton and the photon. The charm quark appears only in the final states of the direct and resolved processes, via the hard scattering of light partons, including the photon, into $c\bar{c}$ pairs. The charm-quark mass is explicitly taken into account together with the transverse momentum of the produced $D^{*+}$ meson as if they were of the same order. In this scheme, the charm-quark mass acts as a cutoff for the initial- and final-state collinear singularities and sets the scale for the perturbative calculations. It is fully retained in the calculation of the hard-scattering cross sections.

Another calculational scheme is the so-called zero-mass variable-flavor-number scheme (ZM-VFNS). This is the conventional parton model approach. In this scheme, the zero-mass parton approximation is applied also to the charm quark, although its mass is certainly larger than $\Lambda_{\text{QCD}}$. Here, the charm quark is also taken into account as an incoming parton with its own parton distribution function (PDF) in the proton and the photon, leading to additional direct and resolved contributions. Usually, charm-quark PDFs are defined with an initial factorization scale $\mu_0$ of the order of $m$. The $D^{*+}$ meson is produced by fragmentation not only from charm quarks produced in the hard scattering processes, but also from the light quarks and the gluon. The transition from partons $u$, $d$, $s$, $c$, $g$ to the $D^{*\pm}$ mesons is described by fragmentation functions (FFs). The well-known factorization theorem provides a straightforward procedure for incorporating these FFs into the order-by-order perturbative calculation. The predictions of this scheme are expected to be reliable only in the region of very large transverse momenta, since terms of the order of $m^2/p_T^2$ present in the hard-scattering cross sections are usually neglected.

A unified scheme that enjoys the virtues of the FFNS and the ZM-VFNS is the so-called general-mass variable-flavor-number scheme (GM-VFNS). In this approach, the large logarithms $\ln(p_T^2/m^2)$ are resummed by the Dokshitzer-Gribov-Lipatov-Altarelli-Parisi (DGLAP) [6] evolution equations for non-perturbative PDFs and FFs, which guarantees the universality of the latter as in the ZM-VFNS, and, at the same time, the mass-dependent terms are retained in the hard-scattering cross section, as in the FFNS.

Unfortunately, the partonic cross sections calculated in the FFNS do not approach their counterparts of the ZM-VFNS in the limit $m \to 0$ (or $p_T \to \infty$), if the collinearly singular terms proportional to $\ln(p_T^2/m^2)$ are subtracted. Therefore, the subtracted FFNS cross sections and the ZM-VFNS cross sections do not approach each other for $m \to 0$, but differ by finite terms. The reason for their occurrence is the different definition of the collinearly singular terms in the two approaches. In the ZM-VFNS calculation, the charm-quark mass is set to zero from the beginning, and the collinearly divergent terms are defined with the help of dimensional regularization. This fixes the finite terms in a specific way in a given factorization scheme, and their form is characteristic to the chosen regularization procedure. If, on the other hand, one starts with $m \neq 0$ and performs the limit $m \to 0$ afterwards, the finite terms are different.

In order to connect the truly massless partonic cross sections of the ZM-VFNS with those of the FFNS, the finite pieces have to be properly subtracted from the latter. The resulting expressions still contain the full mass dependence of the FFNS, but approach their ZM-VFNS counterparts in the limit $m \to 0$. This is the main feature of the GM-VFNS. This approach was applied to $\gamma + \gamma \to D^{*\pm} + X$ [7, 8], $\gamma + p \to D^{*\pm} + X$ [9], and $p + \bar{p} \to D^{*\pm} + X$ [10, 11]. However, the treatment of $\gamma + p \to D^{*\pm} + X$ [9] was not complete, since the resolved contribution was still evaluated in the ZM-VFNS. In the meantime, the results of Refs. [10, 11] have become available, which immediately carry over to resolved photoproduction.

It is the purpose of this work to present results for the full photoproduction cross section where both the direct and resolved contributions are calculated in the GM-VNFS. For the direct part, our calculation is based on Refs. [7, 8], for the resolved part we use the results of Refs. [10, 11]. In addition, we incorporate in our calculation new and better $D^{*\pm}$ FFs in the GM-VFNS [12], which are extracted from recent experimental data of $e^+ + e^- \to D^{*\pm} + X$ from the Belle [13] and CLEO [14] Collaborations. We also study the dependence of the $D^{*\pm}$ photoproduction cross section on the renormalization and factorization scales and on the choice of proton and photon PDFs and of $D^{*\pm}$ FFs, and we also do this separately for the direct and resolved contributions, which was not done in the past. We also investigate the influence of the charm-quark mass in detail as a function of $p_T$ and rapidity $\eta$. In this work, our results will not be compared with the experimental data of the H1 and ZEUS Collaborations, since such comparisons were already shown in a H1 publication [1] and will be in the final H1 publication of the most recent measurements. We also hope to provide detailed numerical results that can be compared with final data from the ZEUS Collaboration; those presented in Ref. [2] are still preliminary.

The outline of this paper is as follows. In Sect. 2, we give a short description of the new $D^{*\pm}$-meson FFs. Sect. 3 contains a detailed discussion of our results, with special emphasis on uncertainties related to the choice of input. A summary is given in Sect. 4.

## 2  $D$-meson fragmentation functions

In Ref. [12], non-perturbative FFs for $D^0$, $D^+$, and $D^{*+}$ mesons were determined by fitting experimental data from the Belle [13], CLEO [14], ALEPH [15], and OPAL [16] Collaborations, taking dominant electroweak corrections due to photonic initial-state radiation into account. These radiative corrections turned out to be significant for the lower-energy data from Belle and CLEO. The fits for $D^0$, $D^+$, and $D^{*+}$ mesons using the Bowler ansatz [17] yielded $\chi^2$/d.o.f. = 4.03, 1.99, and 6.90, respectively. The significance of finite charm- and bottom-mass effects was investigated through comparisons with a similar analysis in the ZM-VFNS. For the conditions of the Belle and CLEO experiments, the effect of taking into account the mass of the $D$ meson in the phase space integration turned out to be appreciable, while charm-quark mass effects on the partonic cross sections were less important. Comparisons of the fit results with the scaled-momentum distributions from Belle and CLEO and the normalized scaled-energy distributions from ALEPH and OPAL were discussed in Ref. [12]. It was found that the Belle and CLEO data tend to drive the average value of the scaled $D$-meson energy $x$, $i.e.$ the scaling variable of the FFs at LO, to larger values, which leads to a worse description of the ALEPH and OPAL data. The FFs resulting from combined fits of Belle, CLEO, ALEPH, and OPAL data, called the Global-GM FFs in Ref. [12], will be used as the default in this work. Their use leads to an improved description of the CDF data of $p + \bar{p} \to D^0, D^+, D^{*+} + X$ [18] from run II at the Tevatron, as may be seen by comparing the relevant figures in Ref. [19, 20] with those in Ref. [21]. In Ref. [12], also fits to the Belle and CLEO data alone were performed yielding $\chi^2$/d.o.f. = 3.13, 1.30, and 3.74 for the $D^0$, $D^+$, and $D^{*+}$ mesons, respectively, $i.e.$ a slightly better description of the data than the Global-GM FFs. The resulting FFs, called Belle/CLEO-GM FFs in Ref. [12], are used as an alternative input to our analysis so as to estimate the theoretical uncertainty from the $D$-meson FFs.

# 3 Results

## 3.1 Comparison of ZM-VFNS and GM-VFNS results

In this subsection, we compare the cross sections of $\gamma+p \to D^{*\pm}+X$ in the ZM-VFNS and GM-VFNS as functions of $p_T$ and $\eta$ in the kinematical regions in which experimental data from H1 [1] and ZEUS [2] exist. As for kinematical cuts, we implement the conditions of the most recent H1 analysis [1] as follows: The energies of the incoming protons and electrons (positrons) are $E_p = 920$ GeV and $E_e = 27.5$ GeV, respectively. The total $\gamma p$ center-of-mass energy $W$ varies in the range $100 \leq W \leq 285$ GeV, which corresponds to an inelasticity $y_e$ in the range $0.1 \leq y_e \leq 0.8$. The maximal value of $Q^2$ allowed by the anti-tagging condition is $Q^2_{\max} = 2$ GeV$^2$. The rapidity is defined in the reference frame of the HERA experiments and taken to be positive in the direction of the incoming proton. The $p_T$ distribution is integrated over $|\eta| \leq 1.5$ and comes as a histogram containing nine bins of varying widths [1]. The $\eta$ distribution is integrated over $1.8 \leq p_T \leq 12.5$ GeV and has six bins of varying widths [1].

As mentioned in Sec. 2, we use the set Global-GM FFs for our default predictions. As further input, we employ the GRV92 photon PDFs [22], converted to the $\overline{\mathrm{MS}}$ scheme, and the CTEQ6.5 proton PDFs [23], which resulted from the first global analysis by the CTEQ group taking into account heavy-quark mass effects with the ACOT$\chi$ prescription. The strong-coupling constant $\alpha_s^{(n_f)}(\mu_R)$ is evaluated from the two-loop formula [24] with $n_f = 4$ active quark flavors and asymptotic scale parameter $\Lambda^{(4)}_{\overline{\mathrm{MS}}} = 328$ MeV corresponding to $\alpha_s^{(5)}(m_Z) = 0.118$, and the charm-quark mass is taken to be $m = 1.5$ GeV. We choose for the renormalization scale $\mu_R$ and the factorization scales $\mu_F$ and $\mu'_F$ of the initial and final states to be $\mu_R = \xi_R m_T$ and $\mu_F = \mu'_F = \xi_F m_T$, where $m_T = \sqrt{m^2 + p_T^2}$ is the transverse mass and $\xi_R$ and $\xi_F$ are dimensionless parameters, which are varied about their default values $\xi_R = \xi_F = 1$ as described in Sec. 3.5 to estimate the scale variation.

We start by discussing the direct contribution to the cross section of $\gamma + p \to D^{*\pm} + X$, which contains finite mass terms. The mass dependence is located in the hard-scattering cross sections for processes with charm in the final state and light quarks and gluons in the initial state. These include the partonic subprocesses $\gamma + g \to c + \bar{c}$ at leading order (LO), where the gluon originates from the proton; virtual corrections to this process combined with the gluon bremsstrahlung subprocess $\gamma + g \to c + \bar{c} + g$; and the NLO subprocess $\gamma + q(\bar{q}) \to c + \bar{c} + q(\bar{q})$, where $q$ denotes a light quark. Explicit expressions for these cross sections and the subtraction terms needed to achieve the transition to the GM-VFNS with $\overline{\mathrm{MS}}$ renormalization and factorization as in the ZM-VFNS can be found for the Abelian part in Ref. [7] and for the non-Abelian part in Ref. [8]. In the non-Abelian part, the finite subtraction terms were calculated by comparing the FFNS calculation by Merebashvili *et al.* [25] with the ZM-VFNS calculation by Gordon [26]. In Ref. [8], two of us found in the calculation of the massless limit of the formulas published in Ref. [25] unexpected subtraction terms $\Delta c_1$, $\Delta \tilde{c}_1$, $\Delta c_2$, and $\Delta c_{11}$ given in Eqs. (43), (45), (47), and (57) of Ref. [8]. These unexpected subtraction terms did not fit into the framework of collinear subtractions of heavy quarks outlined in Ref. [11], according to which all subtraction terms are generated by the convolution of LO partonic cross sections with partonic FFs $d_{Q\to Q}(x,\mu)$ [11]. In this case, the finite subtraction terms are non-zero only in the Abelian part, as was verified in Ref. [7]. It was found out later that such subtraction terms in the non-Abelian part indeed vanish. This mismatch was caused by a misprint in the formula for $F_2$ in Eq. (C2) of Ref. [25]. Actually, these subtraction terms had almost no effect on the results of the previous work [9], since they had been subtracted for the transition to the GM-VFNS. Furthermore, also the subtraction term $\Delta c_{11}$ for $\gamma + q \to c + \bar{c} + q$, given in Eq. (78) in Ref. [8], actually vanishes, as was already explained in our previous work on inclusive $D^{*+}$-meson production in $p\bar{p}$ collisions [10]. Also this term is put to zero in this work.

To study the size of the corrections due to finite mass effects, we first consider the direct contribution to $d\sigma/dp_T$. Its GM-VFNS to ZM-VFNS ratio $d\sigma(m \neq 0)/d\sigma(m = 0)$ is shown as a function of $p_T$ in Fig. 1(a). In the smallest-$p_T$ bin, this ratio is 0.7, but it increases rapidly with increasing values of $p_T$ and reaches 0.96 in the highest-$p_T$ bin. The reduction of the direct cross section due to finite-mass effects is thus significant for $p_T \lesssim 3.5$ GeV. In Fig. 1(a), we also show the GM-VFNS to ZM-VFNS cross section ratio for the sum of the direct and resolved parts, denoted *tot* in the figure. This ratio is almost constant as a function of $p_T$, varying between 0.94 and 0.98 in the considered range of $p_T$ values. The reason for this behavior resides in the fact that the finite-mass corrections decrease the direct part, whereas they increase the resolved part. A similar behavior was observed in our studies of the finite-mass corrections to the hadroproduction of $D^{*+}$ meson in Ref. [10]. Due to the compensation of finite-mass corrections to the direct and resolved components, their overall effect is strongly reduced. We note that the curve labeled *tot* in Fig. 1(a) corresponds to the complete direct contribution with $n_f = 4$ flavors, *i.e.* it includes also the component originating from charm and anticharm quarks in the proton. This component is connected with the factorization of mass singularities in the $g \to c\bar{c}$ channel at the proton vertex and is evaluated in the massless approximation. In Fig. 1(b), we repeat the analysis of Fig. 1(a) for $d\sigma/d\eta$ as a function of $\eta$. Mass effects are more prominent in this case and affect the shape of the distribution. The sum of the direct and resolved parts is decreased by charm-quark mass effects at negative rapidities, by up to $-16\%$ in the leftmost bin, and it is increased for positive rapidities, by about $+50\%$ in the rightmost bin.

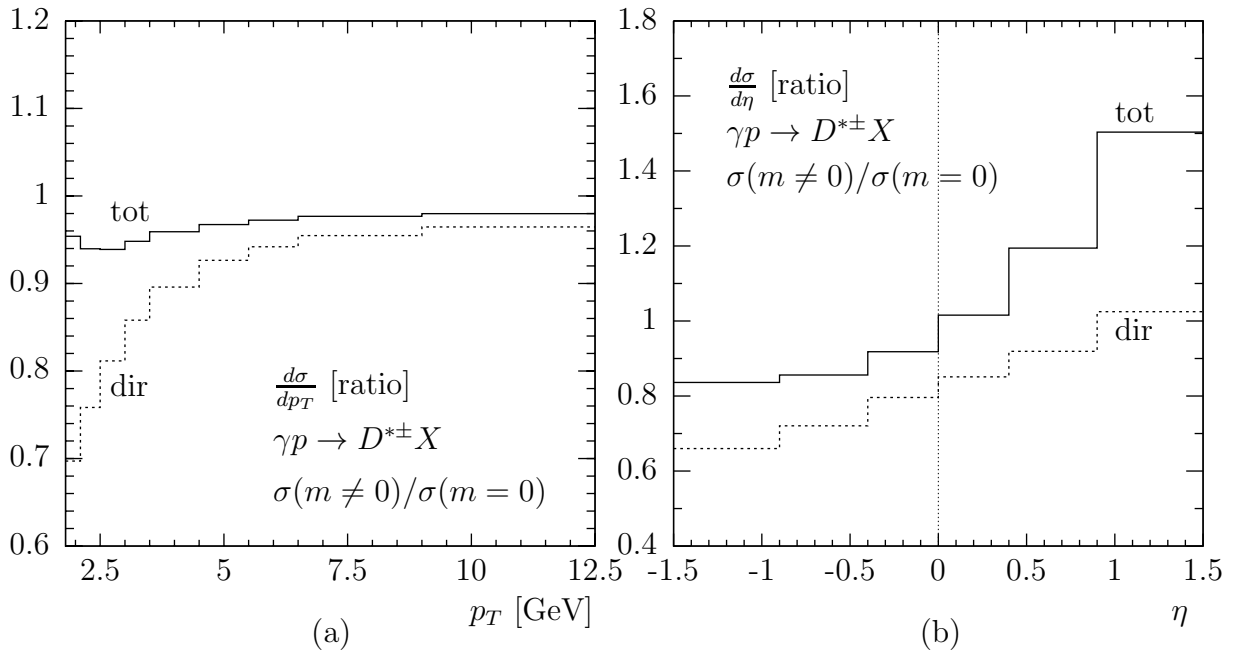

Figure 1: Influence of the finite charm-quark mass on the differential cross sections (a) $d\sigma/dp_T$ and (b) $d\sigma/d\eta$ of $ep \to D^{*\pm} + X$ via photoproduction (tot) and its direct mode (dir) at NLO. The evaluations in the GM-VFNS are normalized to those in the ZM-VFNS.

Mass-dependent terms introduce an additional sensitivity on the value of the charm-quark mass. Varying $m$ by $\pm 0.3$ GeV about its default value of 1.5 GeV, we observe a change of the $p_T$ distribution by a few per cent: the uncertainty is largest at small $p_T$ values, reaching roughly $\pm 6\%$, but decreases to values below $\pm 1\%$ at large $p_T$ values. In the case of the $\eta$ distribution, the uncertainty is $\pm 6\%$ at negative $\eta$ values and decreases towards positive $\eta$ values.

In the calculation described above, we identify the rapidity of the inclusively produced charm quark with the pseudorapidity of the $D^{*+}$ meson, and only the transverse momentum is scaled down when folding with the FF. This corresponds to neglecting the mass of the $D^{*+}$ meson. In fact, there is an inherent ambiguity in the definition of the scaling variable $z$ and the way how to take into account the hadron mass. To estimate the corresponding uncertainties in our predictions, we also adopt a non-trivial definition of the scaling variable $z$, namely as the ratio of the plus-components of the quark and meson four-momenta, *i.e.*, $p_{D^{*+}}^+ = z p_c^+$, where $p^+ = E + p_L$, with $E$ and $p_L$ being the energy and longitudinal momentum in the $\gamma p$ center-or-mass system. The $p_T$ and $\eta$ distributions obtained in this way, with $m_{D^{*+}} = 2.01$ GeV, normalized to their counterparts with zero $D^{*+}$-meson mass are shown in Fig. 2. The ratios are presented separately for the direct (dashed lines) and resolved (dotted lines) contributions, allowing for charm quarks in the initial state. For the complete cross section $d\sigma/dp_T$ shown in Fig. 2(a), the ratio is close to one; except for the first two $p_T$ bins, where we observe an enhancement of up to $+2\%$, we find a small suppression, by less than $-2\%$. However, the shape of the complete $\eta$ distribution $d\sigma/d\eta$ shown in Fig. 2(b) is more strongly affected. At negative rapidities, we find a suppression reaching $-14\%$ in the first $\eta$ bin, and at positive rapidities the cross section is enhanced by up to $+10\%$ in the last $\eta$ bin. Since there is no strong theoretical justification to prefer one over the other prescription to take into account these kinematic mass effects, we neglect this small effect in the following.

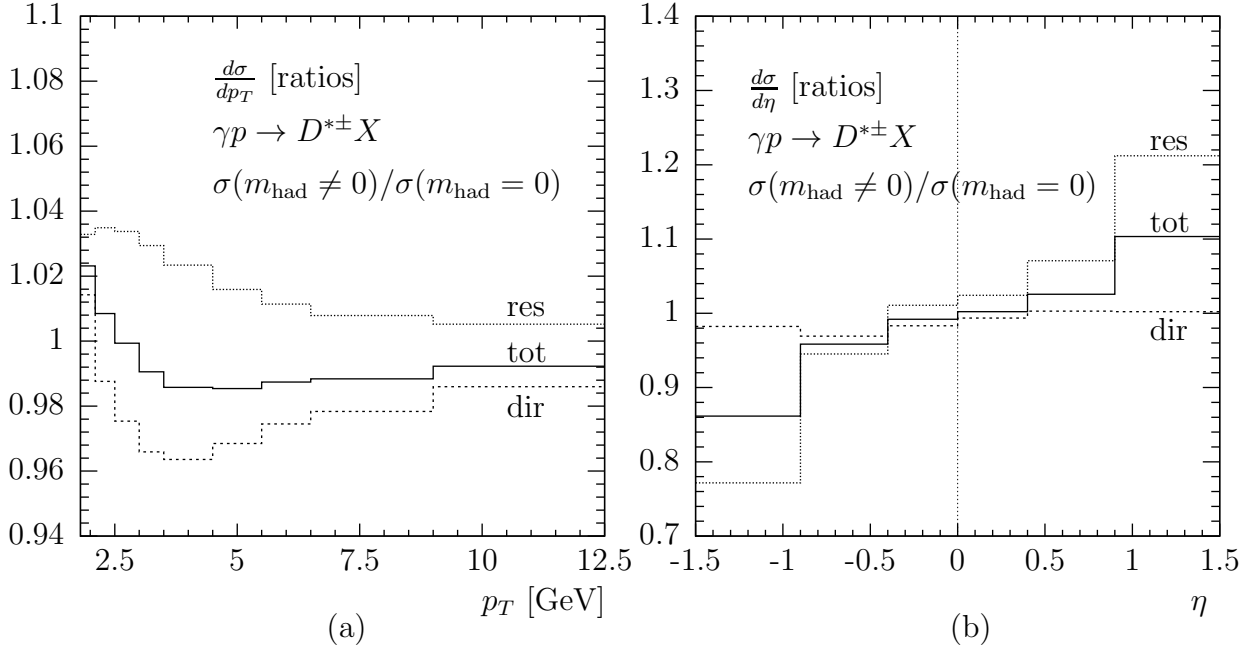

Figure 2: Influence of the finite $D^{*+}$-meson mass $m_{\text{had}}$ on the differential cross sections (a) $d\sigma/dp_T$ and (b) $d\sigma/d\eta$ of $ep \to D^{*\pm} + X$ via direct (dir) and resolved (res) photoproduction and their sum (tot) at NLO in the GM-VFNS. The evaluations with $m_{\text{had}} \neq 0$ are normalized to those with $m_{\text{had}} = 0$. In the case of $m_{\text{had}} \neq 0$, the scaling variable of the FFs is defined as the ratio of the plus-components of the four-momenta of the $D^{*+}$ meson and the fragmenting parton.

### 3.2 Dependence on proton and photon PDFs

The dependence of the photoproduction cross section in the GM-VFNS on the PDFs of the proton has not been investigated in detail in the past. As for the various versions of proton PDFs from CTEQ, we do not expect large variations, since the recent CTEQ parameterizations are very similar to each other. This is also the result of our numerical evaluations shown in Fig. 3(a), where we display, as a functions of $p_T$, the cross sections $d\sigma/dp_T$ evaluated with sets CTEQ5M and CTEQ5M1 [27] and normalized to the result for set CTEQ6.5M, which serves as the default throughout this paper. In fact, we observe from Fig. 3(a) that the cross section ratios for the CTEQ parameterizations are very close to one; the deviations from one are below 4% and largest in the smallest-$p_T$ bin. The picture changes somewhat when we compare with results obtained using the MRST2004 parameterization of the Durham group [28]. In fact, the MRST2004 to CTEQ6.5M ratio for $d\sigma/dp_T$, also shown in Fig. 3(a), is less than one, starting at 0.84 in the smallest-$p_T$ bin and approaching one in the largest-$p_T$ bin. The larger deviation of the cross section with MRST2004 partons is mainly due to the different gluon PDF entering the direct contribution; the different charm PDF is less important and accounts for about a quarter of the observed difference. Similar plots for $d\sigma/d\eta$ are displayed in Fig. 3(b). For all three alternative PDFs discussed above, the ratios to the CTEQ6.5M result are almost independent of $\eta$ in the considered range. The ratios for the CTEQ parameterizations lie very close to one and decrease with increasing value of $\eta$, whereas for MRST2004 the ratio is near 0.9 and increases with increasing value of $\eta$. The most recent PDF set of the CTEQ group, CTEQ6.6M [29], leads to predictions which differ from the results for CTEQ6.5M by less than one per cent over the whole range of $p_T$ and $\eta$ values considered here. In total, we conclude that the dependence on the chosen PDFs is rather weak and still small compared to the scale variation to be examined in Sec. 3.5.

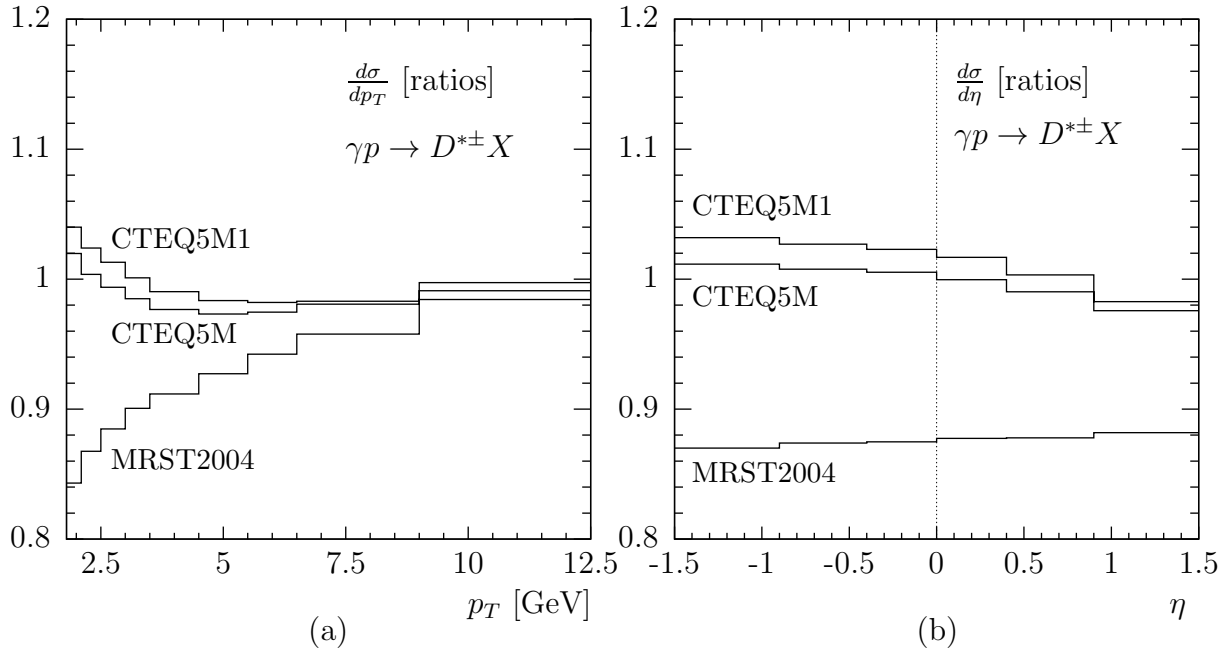

Figure 3: Influence of the proton PDFs on the differential cross sections (a) $d\sigma/dp_T$ and (b) $d\sigma/d\eta$ of $ep \to D^{*\pm} + X$ via photoproduction at NLO in the GM-VFNS. The evaluations with the CTEQ5M, CTEQ5M1 [27], and MRS2004 [28] sets are normalized to those with the CTEQ6.5M [23] set.

A considerable part of the resolved contribution is due to charm quarks in the proton. It is, therefore, interesting to investigate the question whether photoproduction of $D^{*\pm}$ mesons is sensitive to the charm PDF. To this end, we adopt the PDF sets accommodating intrinsic charm recently published by the CTEQ Collaboration [30]. These parameterizations exist in six different versions and are referred to as CTEQ6.5cn, with n = 1, ..., 6. They were obtained and tested by extending the recent CTEQ6.5 global analysis [23] to include charm PDFs with independent parameters fixed at the initial factorization scale $\mu_0 = m$. The amount of intrinsic charm in these six versions was determined in such a way that it is consistent with all the data used in the CTEQ6.5 global analysis. In the latter, the intrinsic charm component is absent and the charm PDF is generated perturbatively via DGLAP evolution. The pairs of versions $n = 1, 2$, $n = 3, 4$, and $n = 5, 6$ differ in the shape of the $x$ distribution of the intrinsic-charm component at the initial scale. The members of each pair differ only slightly in normalization. For details, we refer to Ref. [30]. We calculated the cross section $d\sigma/dp_T$, integrated over $\eta$ in the range $|\eta| \leq 1.5$, as a function of $p_T$ in the range $2 \leq p_T \leq 20$ GeV for all six CTEQ parameterizations with intrinsic charm and normalize the results to the CTEQ6.5 prediction. It turns out that these ratios are very close to one. Specifically, their deviations from one are 0.1–0.3% ($n = 1$), 1% ($n = 2$), 0.3–0.6% ($n = 3$), 1% ($n = 4$), $-0.2$–2% ($n = 5$), and 0.2–4.5% ($n = 6$). The ratios for $n = 2, 4$, and 6 are slightly larger than their counterparts for $n = 1, 3$, and 5. The size of the effect for the different models reflects the properties of the charm distribution built into the various parameterizations by construction. For example, the normalizations of the charm distributions in the models with $n = 2, 4$, and 6 are larger than those for $n = 1, 3$, and 5; and the models with $n = 5$ and 6 have a charm distribution that is enhanced over the whole $x$ range, whereas the models with $n = 1, 2, 3$, and 4 peak at large $x$ values. From this study, we conclude that there is no hope to obtain additional information on the intrinsic charm component in the proton from $D^{*\pm}$ photoproduction data; the observed differences are far too small to be observable. The effect of an intrinsic charm component on $D$-meson production has recently been investigated also for $pp$ and $p\bar{p}$ collisions in Ref. [19], where larger effects, especially for BNL RHIC, were found.

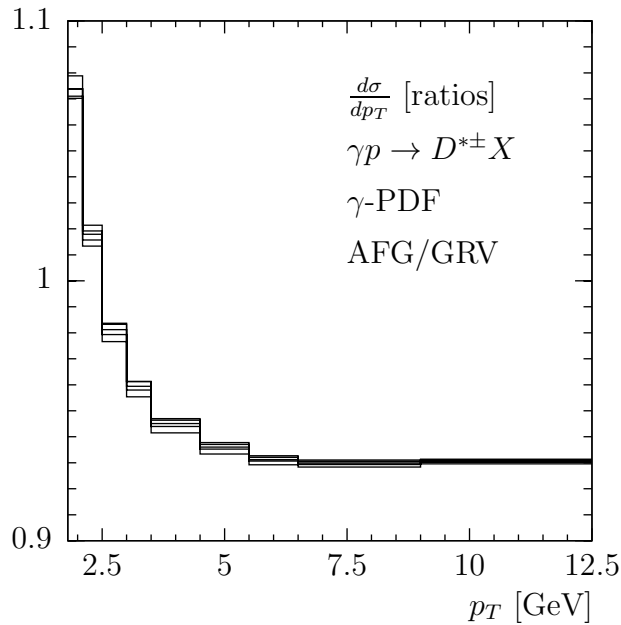

Figure 4: Influence of the photon PDFs on the differential cross section $d\sigma/dp_T$ of $ep \to D^{*\pm} + X$ via photoproduction at NLO in the GM-VFNS. The evaluations with the five AFG [31] sets are normalized to the one with the GRV92 [22] set.

Also the PDFs of the photon are needed to calculate the resolved cross section. The GRV92 NLO set, transformed to the $\overline{\text{MS}}$ scheme, used above serves as our default also in the following. Since 1992, several new photon PDF sets were constructed by fitting to LEP $F_2^\gamma$ data. Here, we consider the most recent NLO photon PDF sets by Aurenche *et al.* [31] denoted AFG04, which come as five versions differing in (i) the choice of initial scale $Q_0^2$, namely $Q_0^2 = 0.34$, 0.70, and 0.97 GeV$^2$; (ii) the normalization of the non-perturbative quark contribution, controlled by the parameter $C_{np}$; and (iii) the hardness of the non-perturbative gluon distribution, controlled by the parameter $p_{10}$. The values of these parameters for the five AFG04 versions, called AFG04$(Q_0^2, C_{np}, p_{10})$, can be found in Table 1 of Ref. [31]. We calculate $d\sigma/dp_T$ for all five parameterizations. The results normalized to the default prediction are plotted as functions of $p_T$ in Fig. 4. We note that all five AFG04 parameterizations yield very similar results, which exceed the default prediction by up to $+8\%$ in the first $p_T$ bin and fall short of it by up to $-7\%$ in the last $p_T$ bin. Although the modern AFG04 PDFs are fitted to a considerably broader set of $F_2^\gamma$ data, in particular from the LEP collaborations, the respective cross sections are remarkably similar to the cross section obtained with the GRV92 parameterization, derived thirteen years earlier. The $\pm 8\%$ difference between the AFG04 and the GRV92 parameterizations is small compared to the scale variation of the cross section to be investigated in Sec. 3.5.

## 3.3 Dependence on FFs

In this subsection, we present results for the cross section distributions $d\sigma/dp_T$, $d\sigma/d\eta$, $d\sigma/dz_D$, where $z_D$ is the fraction of photon energy passed on to the $D^{*+}$ meson in the proton rest frame, and $d\sigma/dW$ for two choices of $D^{*\pm}$ FFs from Ref. [12]. The first set, Global-GM, which is our default choice, was obtained in Ref. [12] by a combined fit to Belle [13] and CLEO [14] data at $\sqrt{s} = 10.52$ GeV and to ALEPH [15] and OPAL [16] data at $\sqrt{s} = m_Z$. The second set, Belle/CLEO-GM, is from a fit to the Belle and CLEO data alone. In Ref. [12], also ZM FF sets were obtained, by neglecting the charm-quark mass in the hard-scattering matrix elements, while taking into account the finite $D^{*+}$-meson mass in the kinematic relations. Since the ZM FFs do not significantly differ from the GM FFs, we do not consider them in this work.

The results for $d\sigma/dp_T$, $d\sigma/d\eta$, $d\sigma/dz_D$, and $d\sigma/dW$ are shown in Figs. 5(a)–(d). $d\sigma/d\eta$, $d\sigma/dz_D$, and $d\sigma/dW$ are integrated over $p_T$ in the region $1.8 \leq p_T \leq 12.5$ GeV, and $d\sigma/dp_T$, $d\sigma/dz_D$, and $d\sigma/dW$ are integrated over $\eta$ in the range $|\eta| \leq 1.5$. The full histograms refer to the Global-GM set and the dotted ones to the Belle/CLEO-GM set. The cross sections evaluated with the latter set exceed those evaluated with the former by 25–30% in average. In the case of $d\sigma/dp_T$, the Belle/CLEO-GM set brings the theoretical prediction into better agreement with the preliminary H1 data [1], which are not shown here, than the Global-GM set, which yields agreement, within errors from scale variation to be discussed in Sec. 3.5, only in the small-$p_T$ region. The results for $d\sigma/d\eta$, $d\sigma/dz_D$, and $d\sigma/dW$ are all dominated by the support from the smallest $p_T$ values, $p_T \gtrsim 1.8$ GeV. Any comparison with experimental data is, therefore, hampered by the large scale variation error, to be discussed in Sec. 3.5. In Figs. 5(b)–(d), the results for the Belle/CLEO-GM FFs again turn out to be slightly larger than those for the default FFs, but the shapes seem to be only feebly affected by the choice of FFs.

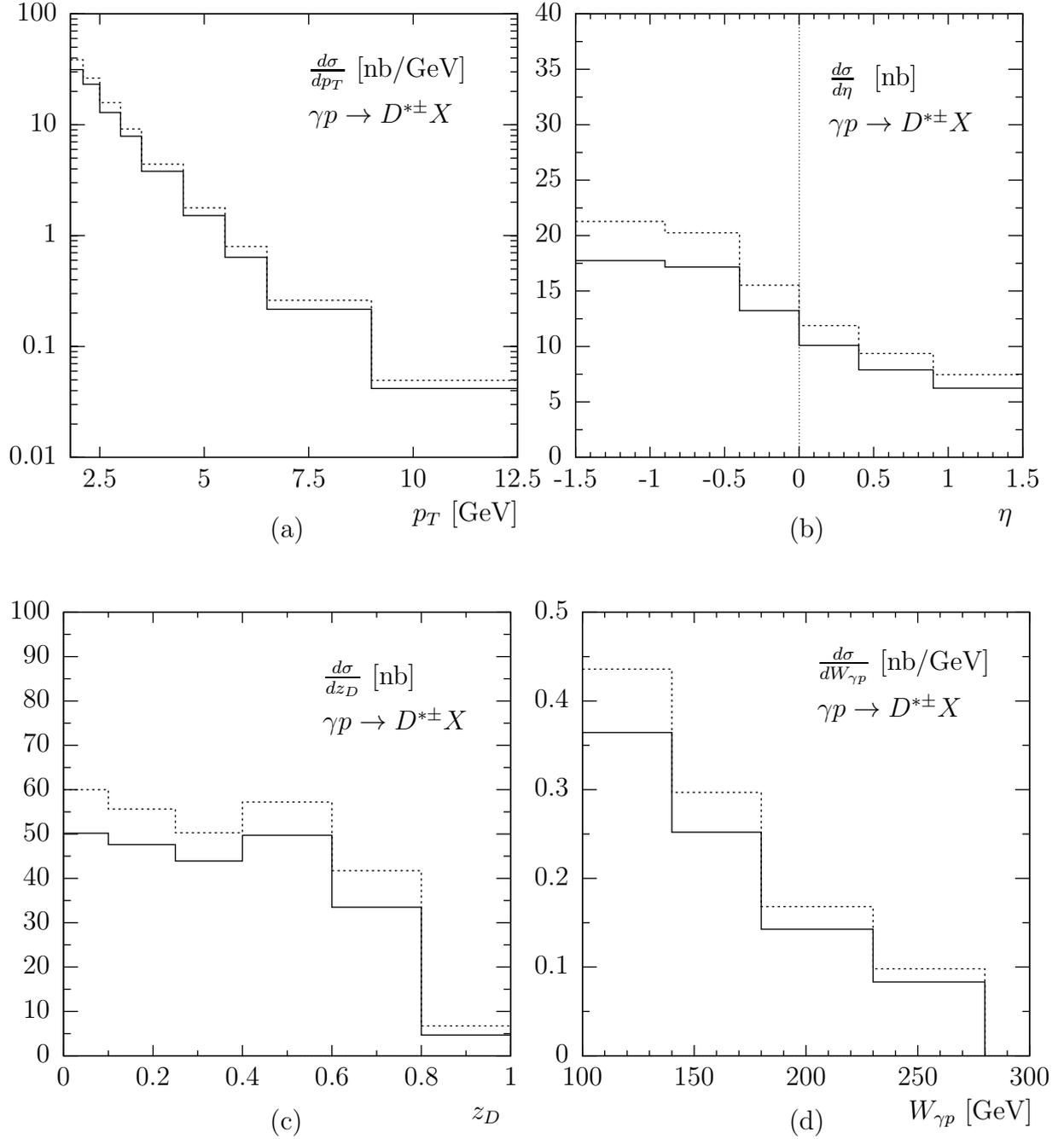

Figure 5: Influence of the $D^{*\pm}$ FFs on the differential cross sections (a) $d\sigma/dp_T$, (b) $d\sigma/d\eta$, (c) $d\sigma/dz_D$, and (e) $d\sigma/dW$ of $ep \to D^{*\pm} + X$ via photoproduction at NLO in the GM-VFNS. The evaluations with the Belle/CLEO-GM (doted lines) and Global-GM (solid lines) sets [12] are compared with each other.

### 3.4 Separation in direct and resolved parts

It is clear that only the resolved part of the cross section depends on the photon PDFs. To understand the rather small dependence of the complete cross section on the choice of photon PDFs observed in Sec. 3.2, it might be interesting to know how the cross section splits up into the direct and resolved parts. Of course, this separation depends on the factorization scheme and scale and is unique only in LO. Therefore, the direct and resolved parts of the cross section are unphysical and cannot be measured separately.

The separation of the complete differential cross section $d\sigma/dp_T$ into its direct and resolved parts as a function of $p_T$ is exhibited in Fig. 6(a). The dashed, dotted, and full histograms represent the direct and resolved parts and their sum, respectively. We observe that the direct part dominates, and that the resolved part decreases somewhat more strongly with increasing value of $p_T$ than the direct part. In the smallest-$p_T$ bin, the resolved and direct parts are nearly equal, whereas in the largest-$p_T$ bin the resolved part amounts to roughly 60% of the direct part. Over the whole $p_T$ range, the resolved part is non-negligible and must be taken into account. It is, to a large extent, due to subprocesses with an incoming charm quark from the proton and, thus, sensitive to the charm PDF of the proton. One should note that also contributions where the charm quark does not take part in the hard-scattering process, but emerges via fragmentation from a light parton, notably the gluon, are non-negligible: at small values of $p_T$ and negative rapidity, the contribution from $g \to D^{*\pm}$ fragmentation reaches the level of 20%.

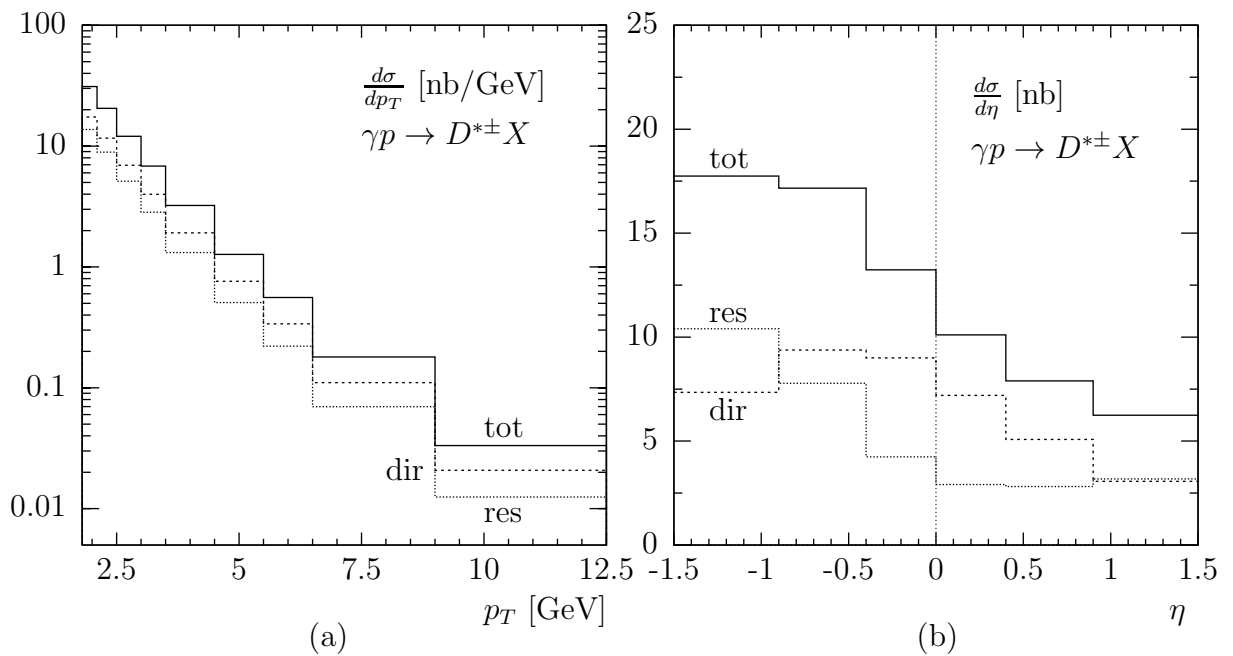

Figure 6: Differential cross sections (a) $d\sigma/dp_T$ and (b) $d\sigma/d\eta$ of $ep \to D^{*\pm} + X$ via direct (dir) and resolved (res) photoproduction and their sum (tot) at NLO in the GM-VFNS.

The separation of the $\eta$ distribution in its direct and resolved contributions is shown in Fig. 6(b). The resolved part is larger than the direct part in the bin $-1.5 \leq \eta \leq -1.0$, *i.e.* in the direction of the incoming quasi-real photon. Here, the resolved part is dominated by the heavy-quark-initiated contribution with charm coming from the photon. In all the other bins, the direct contribution is larger than the resolved one.

## 3.5 Scale dependence

In this subsection, we investigate the scale variation of the differential cross sections $d\sigma/dp_T$ and $d\sigma/d\eta$ in the kinematic ranges described above. In principle, we are dealing with three independent scales, the renormalization scale, $\mu_R$, and the factorization scales, $\mu_F$ and $\mu'_F$, of the initial and final states, respectively. However, we identify the latter two for simplicity and set $\mu_R = \xi_R m_T$ and $\mu_F = \mu'_F = \xi_F m_T$, as already explained in Sec. 3.1. To estimate the scale uncertainty, we independently vary $\xi_R$ and $\xi_F$ in the range $1/2 \leq \xi_R, \xi_F \leq 2$ about their default values $\xi_R = \xi_F = 1$ imposing the constraint $1/2 \leq \xi_F/\xi_R \leq 2$. The results for $d\sigma/dp_T$ obtained with the five choices $(\xi_R, \xi_F) = (\frac{1}{2}, 1)$, $(1, 2)$, $(1, 1)$, $(2, 1)$, and $(1, \frac{1}{2})$ are shown in Fig. 7(a). The cross section is found to be maximal for $(\xi_R, \xi_F) = (\frac{1}{2}, 1)$ and to be minimal for $(\xi_R, \xi_F) = (1, \frac{1}{2})$ if $p_T \lesssim 3$ GeV and for $(\xi_R, \xi_F) = (2, 1)$ otherwise. Outside the small-$p_T$ range, the scale dependence is essentially due to $\xi_R$ alone, as might be expected. The cross section takes its minimal and maximal values for $\xi_F = 1$, except in the small-$p_T$ range, where it becomes minimal for $\xi_F = 1/2$. As expected, the scale variation of $d\sigma/dp_T$ is largest in the smallest-$p_T$ bin, where it reaches $+84\%/-53\%$ of its default value. In the largest-$p_T$ bin, it amounts to merely $+13\%/-16\%$.

The corresponding results for the $\eta$ distribution are shown in Fig. 7(b). Here, the scale variation is large and dominated by the contribution from the lower end of the $p_T$ range $1.8 \leq p_T \leq 12.5$ GeV. In the most negative (positive) $\eta$ bins, the scale change reaches $+76\%/-36\%$ ($+74\%/-46\%$) relative to the default cross section.

## 4 Summary

We presented a detailed discussion of numerical results for the photoproduction of $D^{*\pm}$ mesons, obtained at NLO in the GM-VFNS. This approach combines the virtues of the ZM-VFNS, where large logarithmic corrections are resummed in universal PDFs and FFs, with those of the FFNS, where finite-mass terms are kept. In this paper, we improved our previous work [9] by extending the GM-VFNS approach to the resolved contribution. In addition, new parameterizations of $D^{*\pm}$ FFs were used to obtain numerical predictions of $p_T$ and $\eta$ distributions, which can be compared to experimental data from HERA.

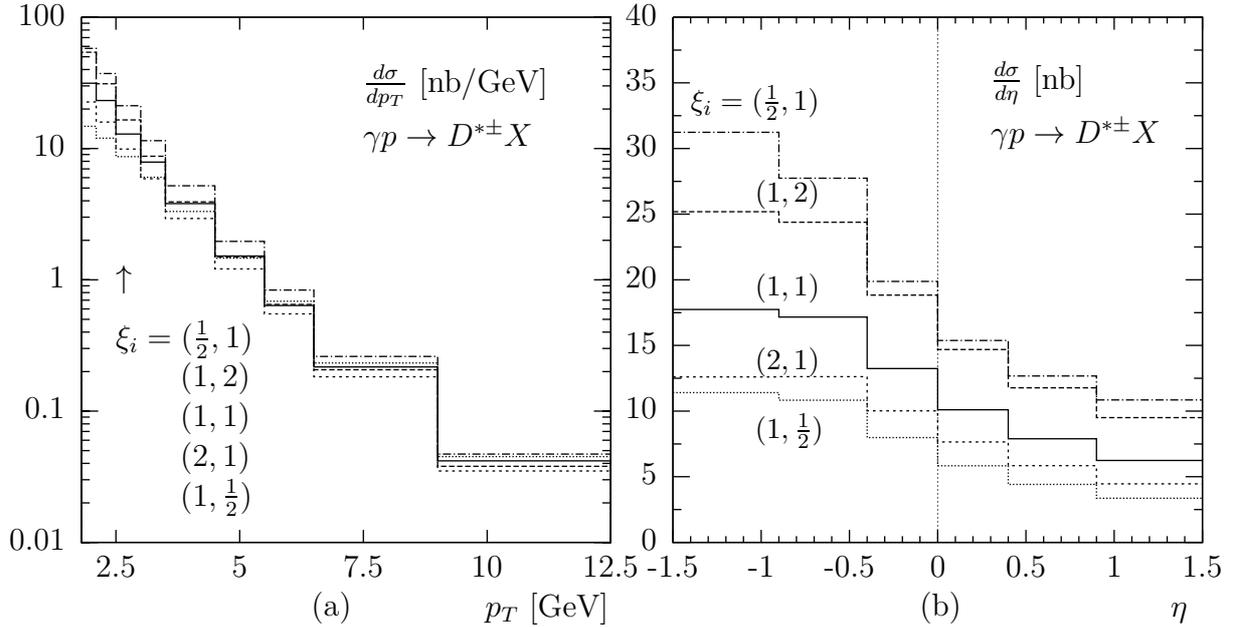

Figure 7: Influence of the renormalization and factorization scales on the differential cross sections (a) $d\sigma/dp_T$ and (b) $d\sigma/d\eta$ of $ep \to D^{*\pm} + X$ via photoproduction at NLO in the GM-VFNS. The evaluations with $(\xi_R, \xi_F) = (\frac{1}{2}, 1)$, $(1, 2)$, $(1, 1)$, $(2, 1)$, and $(1, \frac{1}{2})$ are compared with each other.

Finite-mass effects were found to be important at small values of $p_T$ and for the shape of the $\eta$ distribution, especially for the direct part. The presence of charm-initiated contributions, which are calculated with zero mass, leads to a suppression of mass effects in the complete cross section. Uncertainties due to the value of the charm-quark mass, the PDFs in the proton and photon, and FFs are found to be roughly of the order of 10%, whereas those due to the variations of the renormalization and factorization scales are significant and exceed the errors on present experimental data.

# Acknowledgment

We thank Z. Merebashvili for clarifying a discrepancy in the analytic results of Ref. [25]. This work was supported in part by the German Federal Ministry for Education and Research BMBF through Grant No. 05 HT6GUA, by the German Research Foundation DFG through Grant No. KN 365/7–1, and by the Helmholtz Association HGF through Grant No. Ha 101.